\documentclass[apl,twocolumn,superscriptaddress,reprint]{revtex4}
\usepackage{graphicx,amssymb,amsmath}

\newcommand{\ca}{{Ca$_3$Ru$_2$O$_7$ }}

\begin{document}
\bibliographystyle{apsrev}

\title{Structural and metal-insulator transitions in ionic liquid-gated
Ca$_3$Ru$_2$O$_7$ surface}

\author{Conor P. Puls}
\author{Xinxin Cai}
\author{Yuhe Zhang}

\affiliation{Department of Physics and and Materials Research Institute, Pennsylvania State University, University Park, PA 16802, USA}

\author{Jin Peng}
\author{Zhiqiang Mao}

\affiliation{Department of Physics, Tulane University, New Orleans, LA 70118, USA}

\author{Ying Liu}
\email{liu@phys.psu.edu}

\affiliation{Department of Physics and and Materials Research Institute, Pennsylvania State University, University Park, PA 16802, USA}
\affiliation{Key Laboratory of Artificial Structures and Quantum Control (Ministry of Education), Shanghai Jiao Tong University, 800 Dong Chuan Road, Shanghai 200240, China }

\date{\today}

\begin{abstract}

We report the fabrication and measurements of ionic liquid gated Hall bar
devices prepared on thin \ca flakes exfoliated from bulk single crystals that were grown by
a floating zone method.  
Two types of devices with their electrical transport properties dominated by
$c$-axis transport 
in Type A or that of the in-plane in Type B devices, were prepared. Bulk
physical phenomena, including 
a magnetic transition near 56 K, a structural and metal-insulator transition at
a slightly lower temperature, 
as well as the emergence of a highly unusual metallic state as the temperature
is further lowered, were found 
in both types of devices.  However, the Shubnikov-de Haas oscillations were
found in Type A but not 
Type B devices, most likely due to enhanced disorder on the flake surface.
Finally, the ionic 
liquid gating of a Type B device revealed a shift in critical temperature of
the structural 
and metal-insulator transitions, suggesting that such transitions can be tuned
by the electric field effect.

\end{abstract}

\maketitle

The discovery of odd-parity, spin-triplet superconductivity in
Sr$_2$RuO$_4$\cite{214SC} generated much interest in related compounds in the
Ruddlesden-Popper (R-P) series of (Ca,Sr)$_{n+1}$Ru$_n$O$_{3n+1}$.
Interestingly,
while strontium ruthenates in the R-P series of Sr$_{n+1}$Ru$_n$O$_{3n+1}$
(Sr$_2$RuO$_4$ is the n = 1 member of the series) are all metals, the calcium
ruthenates are
more strongly correlated than their strontium ruthenate counterparts, featuring
metallic as well as insulating behavior accompanied by magnetic, structural, and
metal-insulator phase transitions. In particular, the bilayer calcium ruthenate,
Ca$_3$Ru$_2$O$_7$, features a band-dependent Mott metal-insulator transition at
56 K, followed by a structural as well as metal-insulator transition at 48 K as
the temperature is
lowered.\cite{Cao1997,Ca327Struct,Ca327Mag} Furthermore, a bulk spin valve
behavior featuring colossal magnetoresistance was discovered,
\cite{CaoAnisotropy,AuluckStructSpin2006,Yoshida2004} which was attributed to
the existence of strongly spin-dependent resistive states in
Ca$_3$Ru$_2$O$_7$, which can be tuned by the application of an in-plane field
leading to a
spin-reorientation and a large resistance change.\cite{AuluckStructSpin2006}

Two observations on Ca$_3$Ru$_2$O$_7$ are particularly intriguing. First,
despite of the co-existence of 
the structural and metal-insulator transitions at 48 K indicating strong
coupling among charge, spin, and lattice degrees 
of freedom in Ca$_3$Ru$_2$O$_7$,\cite{OOpredict,327OOtheory} resonant X-ray
scattering measurements 
did not yield any evidence for an orbital ordering in
Ca$_3$Ru$_2$O$_7$,\cite{Bohnenbuck2008} which
raises the question of whether the structural transition is actually
electronically driven. Second, 
a highly unusual metallic state with a very low carrier density was found to
emerge below around 8 K. 
An electronically driven structural transition is a phenomenon of current
technological interest in the context of oxide 
electronics.  Similar phenomena was found in vanadium oxide, which features a
metal-insulator 
transition just above room temperature and has been proposed for
next-generation field-effect 
transistor technologies.\cite{RussiaVO2,HarvardVO2} The emergence of an
unusually low carrier 
density metallic state in an insulating phase, which results in the
Shubnikov-de Haas oscillations (SdHOs), resembles that observed in under doped high $T_c$ superconductors
\cite{OrbTransMetalOx} that was
attributed to the presence of pre-formed electron pairs. As to the metallic
phase found below 8 K,
even though its existence was revealed long ago in the flux grown crystals,\cite{Cao1997,Ca327Struct,Ca327Mag} and confirmed in the flux grown
crystals more 
recently,\cite{MackenzieOscillations} the nature of this phase has rarely been
discussed.
Electric field effect study of this system will provide insight into these
questions.

The challenge of studying the electric field effect of Ca$_3$Ru$_2$O$_7$ is 
two-fold. First, high-quality thin films of Ca$_3$Ru$_2$O$_7$ are difficult to
prepare. 
Furthermore, Ca$_3$Ru$_2$O$_7$ is neither very resistive or electronically
anisotropic, 
making the non-surface contribution to total sample conductance significant for 
any electric field effect samples. The exfoliation of layered materials into
thin 
single-crystal flakes, inspired by the graphene work\cite{Geim2DCrystals} 
provides a solution to the first problem. However, the issue of the small surface
contribution to 
total sample conductance is difficult to address. In this regard, making 
thin flakes and using very high charge density change will help. Specifically, 
significant electric field effect of \ca would require that 10$^{13}$-10$^{15}$
cm$^{-2}$ per
bilayer be achieved.\cite{Ca327Hall} Ionic liquid gating 
techniques, capable of inducing up to 10$^{15}$ cm$^{-2}$
charge carriers\cite{IL1015} by the formation of an electrical double layer
(EDL) at the sample/liquid interface,\cite{ShklovskiiEDL} have previously been
developed for studies of insulating transition metal oxides.  Superconductivity
was discovered in insulating KTaO$_3$ by gating beyond 3 x
10$^{14}$ cm$^{-2}$,\cite{TohokuKTaO3} and in YBCO the superconducting critical
temperature was pushed to zero by depleting a comparable density.\cite{GoldmanYBCO}  EDL gating
has also confirmed carrier-mediation of ferromagnetic ordering in
Ti$_{0.90}$Co$_{0.10}$O$_2$.\cite{TohokuTiCoO2}

\begin{figure}[t!]
\centering
\includegraphics[natwidth=243,natheight=216]{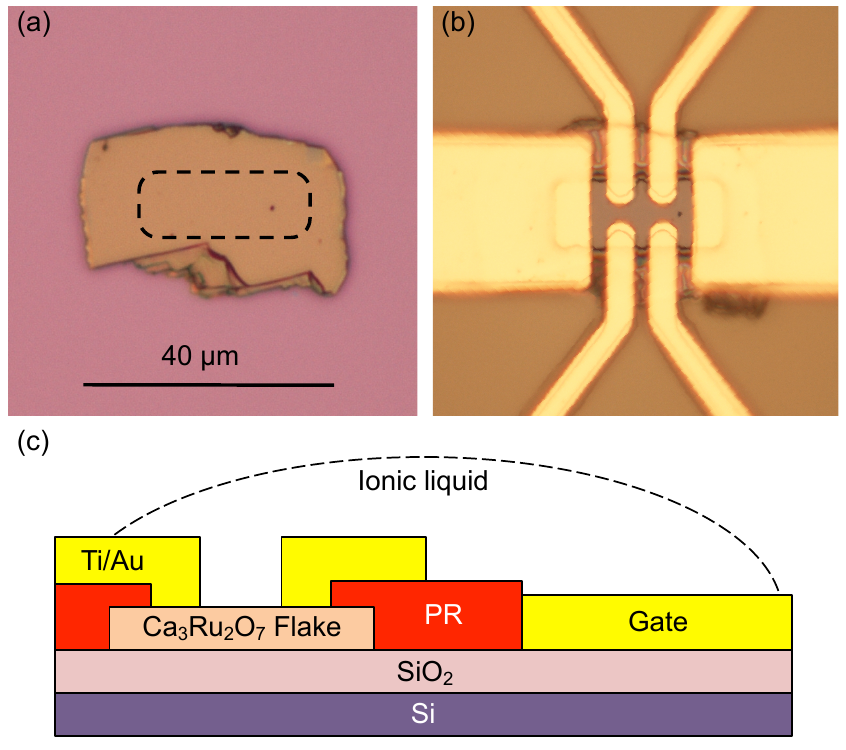}
\caption{(a) Optical image of a \ca flake supported by a Si/SiO$_2$ substrate. 
Dashed lines outline location of photoresist window to be patterned; (b) Optical
image of a device completed on the same flake with electrical contact made on
the surface of the flake using a window through hard-baked photoresist; (c)
Schematic of the side-view of the device, including the coplanar ionic liquid
gate (G) setup, with two of the six contacts acting as source (S) and drain
(D).}
\label{opts}
\end{figure}

Single crystals of \ca used in this study were grown by a floating zone method.
 Flakes 
of \ca were exfoliated via mechanical cleavage from bulk crystals and deposited
onto a
substrate of 300 nm SiO$_2$ thermally grown on undoped Si.  Flakes are
typically on the order of 30-50 $\mu$m in lateral length, and between 0.5 and 1
$\mu$m in thickness along the $c$-axis; one flake is shown in Fig. \ref{opts}a.
 Thickness was estimated by focusing the both the flake and the substrate within
the sub-micron depth of view field of our optical microscope, and confirmed by 
atomic force microscope (AFM) measurements.  We developed a
process to contact only the top surface of the flake by hard-baking a
photo-lithographically defined window on the surface of a flake before defining
metal contacts.  We patterned Ti/Au metal contacts in a Hall bar geometry.  A
short, low-power oxygen etch cleaned the sample surface sufficiently after
processing.  A completed device is shown in Fig. \ref{opts}b.  This
surface-contacted geometry prepares the device for top-gating with an ionic
liquid, shown schematically in Fig. \ref{opts}c, and is preferred to maximize
the surface signal in metallic, though anisotropic, materials. We use the
ionic liquid N,N-diethyl-N-(2-methoxyethyl)-N-methylammonium bis(trifluoromethylsulphonyl-imide), DEME-TFSI) as the gate dielectirc.  Devices were measured within a
Physical Property Measurement System (Quantum Design) with a base temperature
of 1.8 K and a 9T superconducting magnet.  Gate voltage is applied just above
the freezing point of DEME-TFSI at 210 K\cite{JapanDEMETFSI} and the sample is
cooled with the gate voltage held constant. 

\begin{figure}[t!]
\centering
\includegraphics[natwidth=243,natheight=365]{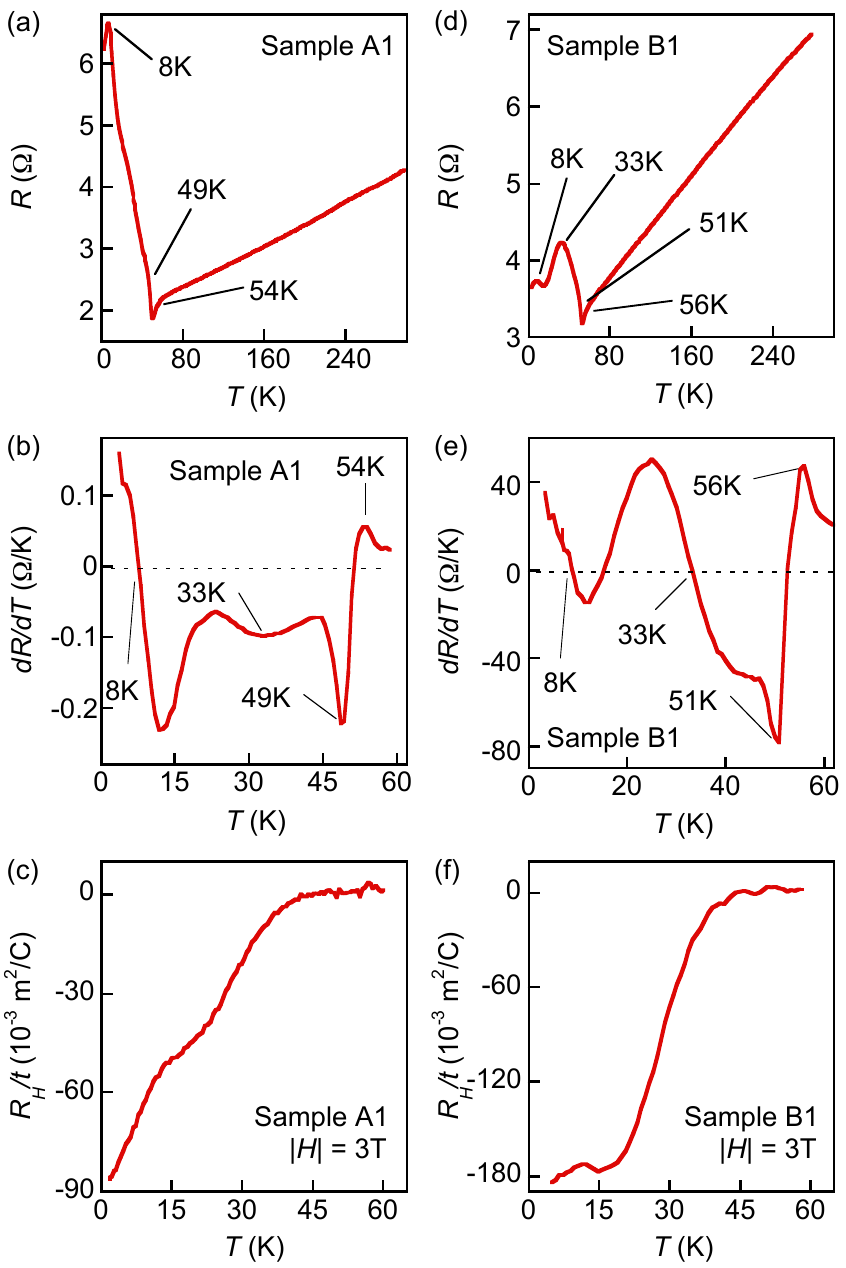}
\caption{(a) Resistance $R$ $vs.$ temperature $T$ in a \ca flake with low-$T$
behavior dominated by $c$-axis transport; (b) $dR/dT$ $vs.$ $T$, calculated
numerically from (a), highlighting complex transition behavior at low
temperatures;  (c) Quotient of the Hall coefficient and thickness $R_H/t$ $vs.$
temperature $T$ measured at $H$ = 3 T in the same device as in (a);
 (d) $R$ $vs.$ $T$ in a \ca flake with low-$T$ behavior dominated
by $ab$-axis transport; (e) $dR/dT$ $vs.$ $T$, calculated numerically from (d); (f) $R_H/t$ $vs.$ $T$ measured at $H$ = 3 T in the same device as in (d).}
\label{RvT}
\end{figure}

In Fig. \ref{RvT}, we show longitudinal resistance $R$ $vs.$ temperature $T$ in
two \ca flake Hall bar devices.  Both devices showed metallic behavior and 
essentially a linear $R \propto T$ behavior until an antiferromagnetic ordering 
transition\cite{Cao1997,Ca327Mag} at 54 and 56 K in these two samples, 
corresponding to the 56 K transition in the bulk, resulting in a sudden drop in 
sample resistance. Lowering temperature further, a structural transition and 
a sharp jump in sample resistance was found near 49 and 51 K, respectively,
corresponding to the 48 K transition in the bulk \cite{Ca327Struct}. However,
qualitatively different behaviors were found for the two devices at low
temperatures.
For Sample A, the insulating behavior was found below the structure transition,
persisting 
to around 8 K, below which a metallic behavior was found. For Sample B, the
insulating behavior lived much shorter than Sample A, with a metallic behavior
found below 33 K. 

The temperature dependence of the sample resistance seen in Sample A is
essentially 
that of the bulk measured along the $c$-axis, $\rho_c$, while that seen Sample
B,  
resembles that of the bulk in-plane resistivity, $\rho_{ab}$.
\cite{MackenzieOscillations, Yoshida2DMetal, Yoshida2004}
The $c$-axis resistivity of the bulk crystals was found to feature more than a
factor of eight 
resistance rise below the structural transition in bulk crystals, and that for our
Sample A is roughly 
factor of 4, which suggests that the sample resistance measured in Sample A
contains 
contributions from both $ab$- and $c$-axis electrical transport. Interestingly,
the temperature
dependence of the sample resistance for Sample A, which we refer here to as a
Type A sample, 
was found in most devices we prepared.  Given that the ratio of $c$-axis
resistivity $\rho_c$ 
to $ab$-axis resistivity $\rho_{ab}$ is only over factor of three, this is not
unexpected. Devices with
behavior resembling to that of Sample B, which we refer to as a Type B sample,
were much 
harder to come by. The sample resistance for Type B samples consists mostly
contribution from the flake 
surface, taking up the behavior of bulk $\rho_{ab}(T)$. 
 
The feature found at 33 K in bulk $\rho_{ab}$\cite{Yoshida2DMetal}, which was
seen in $R$ $vs.$ $T$ for 
Sample B and in $dR/dT$ for Sample A, marks the onset of a quasi
two-dimensional metallic state.
The metallic behavior found in $\rho_c$ below 8 K, on the other hand, signals
a incoherent-coherent 
transition in the $c$-axis transport and the emergence of a fully
three-dimensional metal in 
Ca$_3$Ru$_2$O$_7$. Interestingly, the 8 K feature in $\rho_c$ was observed in
floating zone,\cite{MackenzieOscillations} but not in self-flux grown crystals,\cite{Yoshida2DMetal, Yoshida2004} which
seems
to suggests that coherent $c$-axis transport is fragile, sensitive to disorder.
The observation of the 8 K feature in our Type A
sample therefore suggests that the good crystallinity of our flakes, consistent
with the temperature dependence of the 
Hall coefficient, $R_H$($T$), shown in Fig. 2c. Incidentally, small deviation
from bulk behavior in $R_H$($T$) was found for Sample B 
at low temperatures, likely reflecting the effect of disorder on the flake
surface. 

\begin{figure}[t!]
\centering
\includegraphics[natwidth=243,natheight=243]{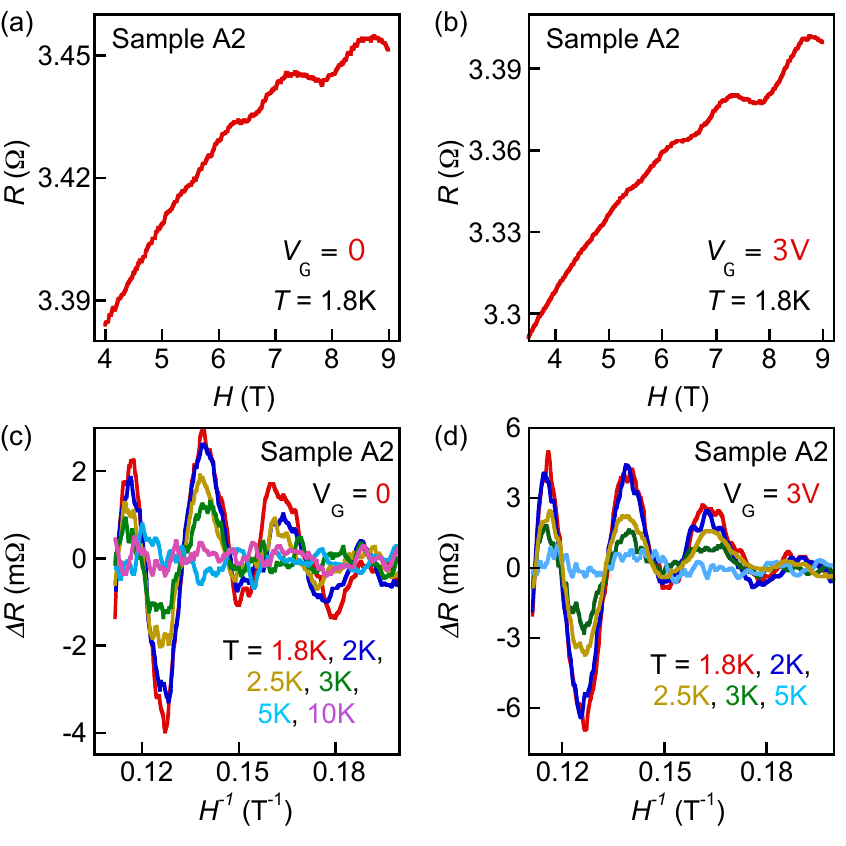}
\caption{(a) Resistance $R$ $vs.$ magnetic field $H$ in a \ca flake with low-$T$
behavior dominated by $c$-axis transport, measured at 1.8K without applying gate voltage;  (b) $R$ $vs.$ $H$ measured at 1.8K with a ionic liquid gating voltage of 3 V applied; (c) and (d) Background-subtracted resistance oscillations $\Delta R$ $vs.$ $H^{-1}$ at different temperatures, numerically calculated from (a) and (b), respectively.}
\label{oscis}
\end{figure}

The above analysis is supported by our observation of SdHOs in Type A
samples which are absent in Type B samples.  We show in 
Fig.\ref{oscis}a $R$ with $H$ at $T$ = 1.8 K in a Type A device. 
The background-subtracted resistance oscillations, $\Delta R$ $vs.$ $H^{-1}$, 
were shown in Fig. \ref{oscis}c.  Similar behavior was found in a separate Type
A device (data not shown). Up to three sets of oscillations were observed in 
bulk \ca.\cite{CaoOsci,MackenzieOscillations} However, a Fourier transform of
$\Delta R$  
obtained in our flake devices suggests only a single set of SdHOs with a
frequency of 43 T, 
which was also seen in the bulk, likely due to a low maximum $H$ in the
present work. 
It is known that the frequency of SdHOs depends on the carrier density. Even
though the 
precise carrier density of the device cannot be obtained from the Hall
measurements because 
the thickness of the layer affected by the gating is not known, the carriers
added to the 
surface can be estimated based on our control experiment carried out on
graphene.
Careful comparison of the SdHOs without gating with those obtained when a ionic
liquid gating 
voltage of 3 V was applied, corresponding a carrier
density change 
larger than 10$^{13}$ cm$^{-2}$, the SdHOs remained essentially the same (Figs.
\ref{oscis}c and d). 
This observation suggests that the SdHOs can not come from the surface of the
flake. 
Incidentally, SdHOs were not observed in Type B samples, which, together with
the deviation 
from the bulk behavior seen in $R_H$($T$), indicates clearly that the transport
in Type B 
samples is dominated by the surface layer featuring disorder stronger than that
in the interior 
of the flake. The surface dominance in Type B samples could be due to
mechanical separation 
formed during the exfoliation process even though there is no direct evidence
for it.

\begin{figure}[t!]
\centering
\includegraphics[natwidth=243,natheight=121]{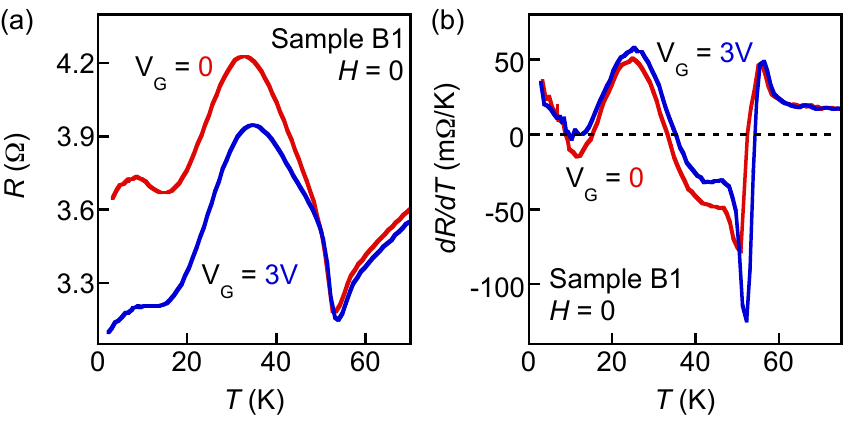}
\caption{(a) Resistance $R$ $vs.$ temperature $T$ in a \ca flake whose
electrical transport is dominated by $ab$-axis transport, for zero and finite
applied gate voltage $V_G$; (b) $dR/dT$ $vs.$ $T$ calculated numerically from
(a).}
\label{gate}
\end{figure}

The surface layer-dominated transport in Type B samples facilitates measurable
response 
to applied gate voltage $V_G$ across an ionic liquid.  In Fig. \ref{gate}a, we
show that 
with $V_G$ = 3 V, the induction of electrons at the flake surface increases
conductivity 
by up to 20\% at lowest temperatures in the metallic regime, likely due to an
added 
carrier density larger than 10$^{13}$ cm$^{-2}$ as mentioned above.  In
numerically 
calculated $dR/dT$ in Fig. \ref{gate}b, we observe a shift in the peak
insulating slope
associated with the structural transition in \ca, from 50 to 53 K. The shift
of a structural
transition with carrier density confirms that the transition is electronically driven
even though orbital ordering is absent. Interestingly, although recent
electrical transport 
studies of bulk \ca under pressure have indicated that the structural
transition is linked to the
long-range antiferromagnetic ordering,\cite{YoshidaPressure} the 56 K
transition observed 
in our Type B sample is barely shifted (Fig. \ref{gate}b).

The three-dimensional metallic state emerged below 8 K is puzzling. The area of 
the primary Fermi surface $\mathcal{A}$ can be estimated from the period in
$H^{-1}$ 
of our SdHOs using the formula $\Delta H^{-1} = 2 \pi e/ \hbar \mathcal{A}$.  A
frequency 
of 43 T gives $\mathcal{A} \approx$ 0.3\% of the 1st Brillouin zone, using
lattice
parameters from Ref. \cite{Ca327Struct}, in agreement with bulk measurements of
both SdHOs and ARPES.\cite{Ca327ARPES} It is intriguing that the onset
temperature for 
this metallic state with a tiny carrier density appears to be unchanged under a
3V ionic 
liquid gating. Together with the fact that the 56 K magnetic transition was
barely shifted by
the same ionic liquid gating of 3 V, our experiment seems to suggest that the
emergence of this 
metallic phase is magnetic in origin.   

In conclusion, we have developed a surface-contact technique for devices
prepared on exfoliated \ca flakes.  Comparison with features seen on these 
devices prepared on exfoliated \ca flakes and those in floating zone-grown 
bulk crystals suggests that the transport properties observed in the Type A and 
Type B samples are dominated by $c$ axis and in-plane contributions,
respectively.  
Magneto electrical transport measurements, including the observation of SdHOs, 
support the emergence of a highly unusual metallic state featuring small Fermi 
surface pockets at low temperatures.  The demonstration of an electric field
effect on 
the structural transition temperature on \ca surface suggests a new approach
to the study of complex transition metal oxides for which thin films are
unavailable.

We would like to thank M. Sigrist, N. Staley and M. Ulrich for useful
discussions. 
The work at Penn State is supported by DOE under Grant No. DE-FG02-04ER46159. The work at Tulane is supported by NSF under DMR-1205469.
The nano fabrication part of this work is supported
by the National Science Foundation (NSF) under Grant
DMR-0908700 and Penn State MRI Nanofabrication
Lab under NSF Cooperative Agreement 0335765, NNIN
with Cornell University. Y. L. also acknowledges support from MOST of China (Grant 2012CB927403) and NSFC
(Grant 11274229) for data analysis and manuscript preparation.

\end{document}